\documentclass[aps,prd,twocolumn,superscriptaddress,showpacs,preprintnumbers,amsmath,amssymb]{revtex4-1}
\usepackage{xspace}
\usepackage{graphicx} 
\usepackage{dcolumn}  
\everymath{\displaystyle}

\newcommand{\gev}{{\rm{\,Ge\kern -0.1em V}}\xspace}
\newcommand{\mev}{{\rm{\,Me\kern -0.1em V}}\xspace}
\newcommand{\gevc}{{{{\rm \,Ge\kern -0.1em V\!/}}c}\xspace}
\newcommand{\mevc}{{{{\rm \,Me\kern -0.1em V\!/}}c}\xspace}
\newcommand{\gevcc}{{{{\rm \,Ge\kern -0.1em V\!/}}c^2}\xspace}
\newcommand{\mevcc}{{{{\rm\,Me\kern -0.1em V\!/}}c^2}\xspace}
\def\KS  {\ensuremath{K^0_{\scriptscriptstyle S}}\xspace}
\def\KL  {\ensuremath{K^0_{\scriptscriptstyle L}}\xspace}
\def\CP {\ensuremath{C\!P}\xspace}
\def\Bbar{\kern 0.18em\overline{\kern -0.18em B}{}\xspace}
\def\Dbar{\kern 0.18em\overline{\kern -0.18em D}{}\xspace}

\graphicspath{{ps}}

\begin{document}



\title{ \quad\\[1.0cm] Measurement of the branching fraction and {$\CP$} asymmetry in {$\boldmath B^{0} \to \pi^{0}\pi^{0}$} decays, and an improved constraint on {$\boldmath\phi_{2}$}  }
\noaffiliation
\affiliation{University of the Basque Country UPV/EHU, 48080 Bilbao}
\affiliation{University of Bonn, 53115 Bonn}
\affiliation{Budker Institute of Nuclear Physics SB RAS, Novosibirsk 630090}
\affiliation{Faculty of Mathematics and Physics, Charles University, 121 16 Prague}
\affiliation{Chonnam National University, Kwangju 660-701}
\affiliation{University of Cincinnati, Cincinnati, Ohio 45221}
\affiliation{Deutsches Elektronen--Synchrotron, 22607 Hamburg}
\affiliation{University of Florida, Gainesville, Florida 32611}
\affiliation{Department of Physics, Fu Jen Catholic University, Taipei 24205}
\affiliation{Justus-Liebig-Universit\"at Gie\ss{}en, 35392 Gie\ss{}en}
\affiliation{SOKENDAI (The Graduate University for Advanced Studies), Hayama 240-0193}
\affiliation{Hanyang University, Seoul 133-791}
\affiliation{University of Hawaii, Honolulu, Hawaii 96822}
\affiliation{High Energy Accelerator Research Organization (KEK), Tsukuba 305-0801}
\affiliation{IKERBASQUE, Basque Foundation for Science, 48013 Bilbao}
\affiliation{Indian Institute of Science Education and Research Mohali, SAS Nagar, 140306}
\affiliation{Indian Institute of Technology Bhubaneswar, Satya Nagar 751007}
\affiliation{Indian Institute of Technology Guwahati, Assam 781039}
\affiliation{Indian Institute of Technology Madras, Chennai 600036}
\affiliation{Indiana University, Bloomington, Indiana 47408}
\affiliation{Institute of High Energy Physics, Chinese Academy of Sciences, Beijing 100049}
\affiliation{Institute of High Energy Physics, Vienna 1050}
\affiliation{Institute for High Energy Physics, Protvino 142281}
\affiliation{INFN - Sezione di Napoli, 80126 Napoli}
\affiliation{INFN - Sezione di Torino, 10125 Torino}
\affiliation{Advanced Science Research Center, Japan Atomic Energy Agency, Naka 319-1195}
\affiliation{J. Stefan Institute, 1000 Ljubljana}
\affiliation{Kanagawa University, Yokohama 221-8686}
\affiliation{Institut f\"ur Experimentelle Kernphysik, Karlsruher Institut f\"ur Technologie, 76131 Karlsruhe}
\affiliation{Kennesaw State University, Kennesaw, Georgia 30144}
\affiliation{King Abdulaziz City for Science and Technology, Riyadh 11442}
\affiliation{Department of Physics, Faculty of Science, King Abdulaziz University, Jeddah 21589}
\affiliation{Korea Institute of Science and Technology Information, Daejeon 305-806}
\affiliation{Korea University, Seoul 136-713}
\affiliation{Kyungpook National University, Daegu 702-701}
\affiliation{\'Ecole Polytechnique F\'ed\'erale de Lausanne (EPFL), Lausanne 1015}
\affiliation{P.N. Lebedev Physical Institute of the Russian Academy of Sciences, Moscow 119991}
\affiliation{Faculty of Mathematics and Physics, University of Ljubljana, 1000 Ljubljana}
\affiliation{Ludwig Maximilians University, 80539 Munich}
\affiliation{University of Malaya, 50603 Kuala Lumpur}
\affiliation{University of Maribor, 2000 Maribor}
\affiliation{Max-Planck-Institut f\"ur Physik, 80805 M\"unchen}
\affiliation{School of Physics, University of Melbourne, Victoria 3010}
\affiliation{University of Miyazaki, Miyazaki 889-2192}
\affiliation{Moscow Physical Engineering Institute, Moscow 115409}
\affiliation{Moscow Institute of Physics and Technology, Moscow Region 141700}
\affiliation{Graduate School of Science, Nagoya University, Nagoya 464-8602}
\affiliation{Kobayashi-Maskawa Institute, Nagoya University, Nagoya 464-8602}
\affiliation{Nara Women's University, Nara 630-8506}
\affiliation{National Central University, Chung-li 32054}
\affiliation{National United University, Miao Li 36003}
\affiliation{Department of Physics, National Taiwan University, Taipei 10617}
\affiliation{H. Niewodniczanski Institute of Nuclear Physics, Krakow 31-342}
\affiliation{Nippon Dental University, Niigata 951-8580}
\affiliation{Niigata University, Niigata 950-2181}
\affiliation{Novosibirsk State University, Novosibirsk 630090}
\affiliation{Osaka City University, Osaka 558-8585}
\affiliation{Pacific Northwest National Laboratory, Richland, Washington 99352}
\affiliation{University of Pittsburgh, Pittsburgh, Pennsylvania 15260}
\affiliation{University of Science and Technology of China, Hefei 230026}
\affiliation{Soongsil University, Seoul 156-743}
\affiliation{Stefan Meyer Institute for Subatomic Physics, Vienna 1090}
\affiliation{Sungkyunkwan University, Suwon 440-746}
\affiliation{School of Physics, University of Sydney, New South Wales 2006}
\affiliation{Department of Physics, Faculty of Science, University of Tabuk, Tabuk 71451}
\affiliation{Tata Institute of Fundamental Research, Mumbai 400005}
\affiliation{Toho University, Funabashi 274-8510}
\affiliation{Department of Physics, Tohoku University, Sendai 980-8578}
\affiliation{Earthquake Research Institute, University of Tokyo, Tokyo 113-0032}
\affiliation{Department of Physics, University of Tokyo, Tokyo 113-0033}
\affiliation{Tokyo Institute of Technology, Tokyo 152-8550}
\affiliation{Tokyo Metropolitan University, Tokyo 192-0397}
\affiliation{University of Torino, 10124 Torino}
\affiliation{Virginia Polytechnic Institute and State University, Blacksburg, Virginia 24061}
\affiliation{Wayne State University, Detroit, Michigan 48202}
\affiliation{Yamagata University, Yamagata 990-8560}
\affiliation{Yonsei University, Seoul 120-749}
  \author{T.~Julius}\affiliation{School of Physics, University of Melbourne, Victoria 3010} 
    \author{M.~E.~Sevior}\affiliation{School of Physics, University of Melbourne, Victoria 3010} 
  \author{G.~B.~Mohanty}\affiliation{Tata Institute of Fundamental Research, Mumbai 400005} 
  \author{I.~Adachi}\affiliation{High Energy Accelerator Research Organization (KEK), Tsukuba 305-0801}\affiliation{SOKENDAI (The Graduate University for Advanced Studies), Hayama 240-0193} 
  \author{H.~Aihara}\affiliation{Department of Physics, University of Tokyo, Tokyo 113-0033} 
  \author{S.~Al~Said}\affiliation{Department of Physics, Faculty of Science, University of Tabuk, Tabuk 71451}\affiliation{Department of Physics, Faculty of Science, King Abdulaziz University, Jeddah 21589} 
  \author{D.~M.~Asner}\affiliation{Pacific Northwest National Laboratory, Richland, Washington 99352} 
  \author{V.~Aulchenko}\affiliation{Budker Institute of Nuclear Physics SB RAS, Novosibirsk 630090}\affiliation{Novosibirsk State University, Novosibirsk 630090} 
  \author{T.~Aushev}\affiliation{Moscow Institute of Physics and Technology, Moscow Region 141700} 
  \author{R.~Ayad}\affiliation{Department of Physics, Faculty of Science, University of Tabuk, Tabuk 71451} 
% ...
\collaboration{The Belle Collaboration}


\begin{abstract}
We measure the branching fraction and $\CP$ violation asymmetry in the decay $B^{0}\to \pi^{0}\pi^{0}$, using a data sample of $752\times 10^{6}$ $B\Bbar$ pairs collected at the $\Upsilon(4S)$ resonance with the Belle detector at the KEKB $e^{+}e^{-}$ collider. The obtained branching fraction and direct $\CP$ asymmetry are $  \mathcal{B}(B\to \pi^{0}\pi^{0}) = [1.31 \pm 0.19~ \text{(stat.)} \pm 0.19~ \text{(syst.)}] \times 10^{-6}$ and $
 A_{\CP}  =  +0.14 \pm 0.36~ \text{(stat.)} \pm 0.10~ \text{(syst.)}, $  respectively. The signal significance, including the systematic uncertainty, is 6.4 standard deviations. We combine these results with Belle's earlier measurements of $B^{0}\to \pi^{+} \pi^{-}$ and $B^{\pm} \to \pi^{\pm} \pi^{0}$ to exclude the $\CP$-violating parameter $\phi_{2}$ from the range $15.5^{\circ} < \phi_{2} < 75.0^{\circ}$ at 95\% confidence level.

\end{abstract}


\maketitle

\tighten

{\renewcommand{\thefootnote}{\fnsymbol{footnote}}}
\setcounter{footnote}{0}

Extensive studies by the Belle, BaBar and LHCb experiments~\cite{PBF,LHCb_CPV_phi1,LHCb_CPV_phi3} have shown that the $\CP$ violation observed in nature can be attributed to a single irreducible phase in the Cabibbo-Kobayashi-Maskawa (CKM) matrix, as proposed by Kobayashi and Maskawa~\cite{KM}.
The unitarity constraint of the CKM matrix, when applied to $B$ mesons and plotted in the complex plane, results in a triangle with internal angles $\phi_{1}$, $\phi_{2}$, and $\phi_{3}$~\cite{alpha}.
Nonzero values for these angles imply $\CP$ violation in the $B$ meson system.
A main objective of the aforementioned experiments is to overconstrain the unitary triangle in order to precisely test the KM mechanism for $\CP$ violation as well as to search for new physics effects. 

One of the proposed techniques to measure $\phi_{2}$ is to perform an isospin analysis of the entire $\pi\pi$ system~\cite{GronauLondon}. This requires measurements of branching fraction ($\mathcal{B}$) and time-dependent $\CP$ asymmetry for the $B^{0} \to \pi^{+}\pi^{-}$ decay, for which Belle recently published precise measurements \cite{Dalseno}, together with measurements of $\mathcal{B}$ and the direct $\CP$ asymmetry ($A_{\CP}$) for $B^{+} \to \pi^{+} \pi^{0}$ and $B^{0} \to \pi^{0} \pi^{0}$ decays~\cite{CC}.
Measurements of all these observables are required as electroweak tree and loop processes contribute with different phases to $B \to \pi \pi$ decays and their effects must be disentangled to determine $\phi_{2}$.
Among the $B \to \pi \pi$ decays, $\mathcal{B}$ and $A_{\CP}$ for $B^{0} \to \pi^{0} \pi^{0}$ are the least well determined.
This decay is also important to probe the disagreement between quantum-chromodynamics-based factorization, which predicts $\mathcal{B}$ below ∼ $1 \times 10^{-6}$~\cite{LiMishima1, LiMishima2}, and previous measurements from Belle and BaBar of (1.8 -- 2.3) $\times 10^{-6}$~\cite{Belle_pi0pi0, Babar_hh}.

In this paper, we present new measurements of $B^{0}\to\pi^{0}\pi^{0}$ based on a 693 fb$^{-1}$ data sample that
contains $752 \times 10^6 B\Bbar$ pairs, collected  with the Belle detector at the KEKB asymmetric-energy $e^+e^-$ ($3.5$ on $8.0\gev$) collider~\cite{KEKB} operating near the $\Upsilon(4S)$ resonance.
In addition, we employ an 83.5 fb$^{-1}$ data sample recorded from runs where the center-of-mass (CM) energy was $60\mev$ below the $\Upsilon(4S)$ resonance (off-resonance data) to characterize backgrounds.

The Belle detector~\cite{Belle} is a large-solid-angle magnetic spectrometer that consists of a silicon vertex detector (SVD), a 50-layer central drift chamber (CDC), an array of aerogel threshold Cherenkov counters, a barrel-like arrangement of time-of-flight scintillation counters, and an electromagnetic calorimeter (ECL) consisting of CsI(Tl) crystals.
All these detector components are located inside a superconducting solenoid coil that provides a 1.5\,T magnetic field.
An iron flux-return located outside of the coil is instrumented with resistive plate chambers to detect $\KL$ mesons and to identify muons.
Two inner detector configurations were used: A 2.0 cm beam-pipe
and a 3-layer SVD were used for the first sample of $132 \times 10^6$ $B\Bbar$ pairs (SVD1), while a 1.5 cm beam-pipe, a 4-layer SVD, and a small-cell CDC were used to record the remaining $620 \times 10^6$ $B\Bbar$ pairs (SVD2)~\cite{svd2}. 

We reconstruct  $B^{0}\to\pi^{0}\pi^{0}$ candidates from the subsequent  decay of $\pi^{0}$ mesons to two photons.
In addition to photons reconstructed from ECL clusters, which do not match any charged track in the CDC, photons that convert to $e^{+}e^{-}$ pairs in the SVD are recovered and reconstructed as
$\pi^{0} \to \gamma e^{+} e^{-}$. This provides a $5.3\%$ increase in detection efficiency.
The photons must have an energy greater than $50$ $(100) \mev$ in the barrel (endcap) region of the ECL.
The invariant mass of the two-photon combination must lie in the range $115\mevcc < m_{\gamma \gamma} < 152\mevcc$, corresponding to $\pm 2.6 \sigma$ around the nominal $\pi^{0}$ mass, and must have a reasonable mass-constrained fit. 

Two kinematic variables are used to distinguish signal from background: the beam-energy-constrained mass, $M_{\text{bc}} \equiv \sqrt{E_{\text{beam}}^{2} - |\vec{p}_{B}|^{2}c^{2}}$, and the energy difference $\Delta E \equiv E_{B} - E_{\text{beam}}$.
Here, $\vec{p}_{B}$ and $E_{B}$ are the momentum and energy of the $B$-meson candidates in the CM frame, and $E_{\text{beam}}$ is half the CM energy of the $e^{+}e^{-}$ collision.
All candidates satisfying $M_{\text{bc}} > 5.26\gevcc$ and $-0.3\gev < \Delta E < 0.2\gev$ are retained for further analysis.
We find that $7.2\%$ of events have more than one $B^{0}$ candidate.
In those cases, we choose the candidate that has the minimum deviation of the two $\pi^{0}$'s reconstructed invariant masses from the world average~\cite{PDG_chi_d}.
This is $90\%$ efficient at selecting the correct $B^{0}$.

The largest background arises from the $e^{+} e^{-} \to q \bar{q} \; (q \in \{u, d, s, c\})$ continuum events.
To suppress this, we construct a Fisher discriminant from 16 modified Fox-Wolfram moments \cite{SFW}. 
To further improve the distinguishing power, we combine the output of the Fisher discriminant with the cosine of the polar angle of the $B$ candidate with respect to the $z$-axis, which is opposite the direction of the $e^{+}$ beam, along with the cosine of the angle between the thrust axis of the $B$ candidate and rest of event in the CM frame.
This creates a final Fisher discriminant ($T_c$) with value in the range ($-1, +1$).
The values near $-1$ ($+1$) denote events having strong continuum ($B$-decay) characteristics.
All candidates with $T_c$ values below $-0.3$ are discarded, removing $72\%$ of the continuum background while retaining $98\%$ of signal events. We subsequently use $T_c$ as a fit variable.

Monte Carlo (MC) simulation studies~\cite{evtgen,GEANT} show that background events that arise from $b \to c$ transitions are mostly due to out-of-time events originating from $e^{+}e^{-}$ interactions such as Bhabha scattering, which leave large energy deposits in the ECL.
Due to the finite decay-time of the CsI(Tl) scintillation, significant residual light from such interactions could still be present in the ECL when a subsequent genuine hadronic interaction occurs.
This ``pileup'' event resembles a hadronic event with high energy back-to-back photons in the CM frame, and thus passes the first-level trigger. 
When combined with random photons from the hadronic interaction, they appear as two $\pi^{0}$'s with a large invariant mass.
Because the energy deposits are almost back-to-back in the CM system, their momentum sum is close to zero which causes the events to peak near the nominal $B$ mass in $M_{\text{bc}}$.
Since the events are recorded in coincidence with hadronic interactions, they also mimic $B$-like events in the continuum suppression variable $T_{c}$. 
A criterion on the trigger time of the CsI(Tl) crystals, which selects ECL interactions in-time with the rest of the event, is employed to suppress this background.
Studies of a high-statistics control mode $B^{0} \to \Dbar^{0}(K^{+}\pi^{-}\pi^{0}) \pi^{0}$, containing 1600 events, show that this requirement removes $99\%$ of the pileup background at the cost of only $1\%$ of signal. 
After applying the timing criterion, we find no background contribution due to $b \to c$ transitions. 
The ECL timing information was initially missing in the SVD1 data set, but was recovered in a subsequent reprocessing of the available raw data.

Other sources of background are found from a dedicated study of rare $B$ decays proceeding via $b\to u,d,s$ transitions in an MC sample 50 times larger than that expected in the recorded data.
The largest of these is due to $B^{+} \to \rho^{+} \pi^{0}$, where the charged pion from the subsequent $\rho^{+} \to \pi^{+} \pi^{0}$ decay is lost.
This background peaks at similar values of $M_{\text{bc}}$ and $T_{c}$ as the signal, but has $\Delta E$ shifted to negative values due to energy loss from the missing $\pi^{+}$.
All other such rare background events (many originate from $B \to \KS(\pi^{0} \pi^{0}) \pi^{0}$, where one of the $\pi^{0}$'s is lost), are shifted to even more negative values in $\Delta E$.
We denote these background events ``rare'' in subsequent text.

The direct $\CP$ violation parameter, $A_{\CP}$, for the $B \to \pi^{0} \pi^{0}$  decay is defined as:
\begin{equation} \label{eq:acp}
  A_{\CP} = \frac{\Gamma(\Bbar^{0}\to \pi^{0} \pi^{0}) - \Gamma(B^{0}\to \pi^{0} \pi^{0})}{\Gamma(\Bbar^{0}\to \pi^{0} \pi^{0}) + \Gamma(B^{0}\to \pi^{0} \pi^{0})},
\end{equation}
where $\Gamma$ is the partial decay width for the corresponding decay.
To measure $A_{\CP}$, we must determine what fraction of the observed $B \to \pi^{0} \pi^{0}$ events originate from $\Bbar^{0}$ or $B^{0}$.
The $B^{0}\Bbar^{0}$ pair originating from the $\Upsilon(4S)$ are produced in a coherent quantum-mechanical state, from which one meson ($B^{0}_{\rm rec}$) may be reconstructed in the $B \to\pi^{0}\pi^{0}$ decay mode.
The $b$-flavor of the other $B$ meson ($B^{0}_{\rm tag}$) can be identified using information from the remaining charged particles and photons.
This dictates the flavor of $B^{0}_{\rm rec}$ as it must be opposite that of the $B^{0}_{\rm tag}$ at
the time $B^{0}_{\rm tag}$ decays.
We follow the procedure described in Ref.~\cite{kukuno} to the determine the $b$-flavor of  $B^{0}_{\rm tag}$.
The tagging information is given by two parameters: the $b$-flavor charge $q$ [$+1$ ($-1$) tagging a $B^{0}$ ($\Bbar^{0}$)], and the purity for flavor charge $r$. 
The value of $r$ is continuous and determined on an event-by-event basis with an algorithm trained on MC events, ranging from zero for no flavor discrimination to one for an unambiguous flavor assignment.
To obtain a data-driven value for $r$, we divide its range into seven regions and determine the mistagging probability, $w$, for each region using a control sample~\cite{kukuno}.
The $\CP$ asymmetry in data is thus diluted by the factor ($1 - 2w$).
Since we do not determine the time between $B^{0}_{\rm rec}$ and $B^{0}_{\rm tag}$,  there is an additional dilution due to $B\Bbar$ mixing, which is accounted for by a factor ($1 - 2\chi_{d}$), with $\chi_{d} = 0.1875 \pm 0.0017$~\cite{PDG_chi_d} being the time-integrated $B\Bbar$-mixing parameter.

The signal yield and $A_{\CP}$ are extracted via an unbinned extended maximum likelihood fit to the four categories of events described by probability density functions (PDFs).
These categories comprise the $B\to \pi^{0}\pi^{0}$ signal described by the $P^{s}$ PDF, continuum ($P^{c}$ PDF), $\rho^{+} \pi^{0}$ ($P^{\rho\pi}$ PDF), and other rare $B$-decay ($P^{r}$ PDF) backgrounds.
Separate PDFs are constructed for the SVD1 (S1) and SVD2 (S2) data sets.
We divide the data into seven bins each for positive and negative $q$-tagged $r$-values for both S1 and S2.
The signal yield and $A_{\CP}$ are determined via a simultaneous fit to the subsequent 28 data sets in three dimensions: $M_{\text{bc}}$, $\Delta E$, and $T_{c}$.

The total likelihood for the 17\,270 events selected as $B^{0}\to\pi^{0}\pi^{0}$ candidates in the fit region is given by
\begin{align}
\label{eq:likelihood}
  {\cal L} & =  \frac{e^{-\sum_{x}N^{x}}}{ \prod_{i,d} N_{i,d}!}
  \times \prod_{i,d} \nonumber \\
  & \left[ \prod_{j=1}^{N_{i,d}} 
   \left( \sum_{x}f^{x}_{i,d}N^{x}P^{x}_{i,d}\left(M_{\text{bc}}^{j},\Delta E^{j},T_{c}^{j},q^{j}\right) \right) \right],
\end{align}
where  $N_{i,d}$ is the number of events in the $i^{\rm th}$ $q\cdot r$ bin for the data set $d$ ($d \in {\textrm{S1}, \textrm{S2}}$) and $N^{x}$ is the number of events in the $x^{\rm th}$ category ($x \in {s,c,\rho \pi, r}$), contributing to the total yield.
The fraction of events in each $i^{\rm th}$ bin for the data set $d$ and $x^{\rm th}$ category is $f^x_{i,d}$ with $\sum_{i,d} f^{x}_{i,d} = 1$.
These fractions implicitly include a factor of half due to the division of the data into positive and negative bins in $q$.
$P^x_{i,d}$ is the three-dimensional PDF for the $x^{\rm th}$ category and $i^{\rm th}$ $q \cdot r$ bin in the $d$ data set, measured at $M_{\text{bc}}^{j}$, $\Delta E^{j}$ and $T_c^{j}$ for the $j^{\rm th}$ event.

The PDF for the signal component is given by:
\begin{align} \label{eq:pdf}
  P^{s}_{i,d}(M_{\text{bc}},\Delta E,T_{c},q) & = [1 - q\times \Delta w_{i,d} + \nonumber \\  & q(1-2w_{i,d}) 
    \times (1-2\chi_{d})A_{\CP}] \nonumber \\
  & \quad P^{s}(M_{\text{bc}},\Delta E,T_{c}),
\end{align}
where $q$ is determined for the $i^{\rm th}$ bin of the data set.
The model takes account of direct $\CP$ violation asymmetry, $A_{\CP}$, and the fractions of signal and backgrounds expected in each combination of S1 (S2) and bin in $q\cdot r$.
In Eq.~\eqref{eq:pdf}, $ \chi_{d}$ is the $B^{0}$ mixing parameter, $w_{i,d}$ is the wrong-tag fraction, and $\Delta w_{i,d}$ is the difference in wrong tag fraction between positive and negative $b$-flavor tags for bin $i$ and data set $d$.
The parameters $w_{i,d}$, $f^{s}_{i,d}$ and  $\Delta w_{i,d}$ are obtained via an analysis of flavor-specific final states using the method described in Ref.~\cite{kukuno}.
The parameters $f^{\rho \pi}_{i,d}$ and  $f^{r}_{i,d}$ are set equal to  $f^{s}_{i,d}$.
The systematic uncertainty arising from this assumption is included in the measurement.

The fraction of continuum events in bin $i$ and data set $d$, $f^{c}_{i,d}$, is determined from fits to off-resonance data.
The ratio of $f^{s}_{i,d}$ for S1 and S2 is fixed to the value expected from the luminosity and detection efficiency.
We determine $N^{\rho \pi}$ and $N^{r}$ from the combination of detection efficiency and expected $\mathcal{B}$.
These are fixed during the fit.
The systematic uncertainties resulting from these assumptions are included in the measurement.
The number of signal events $N^{s}$, asymmetry $A_{\CP}$, the number of continuum events $N^{c}$, and the ratio between the total number of continuum events in S1 and S2 are free parameters in the fit to the data.

In the case of signal, there is a significant correlation between $M_{\text{bc}}$ and $\Delta E$ due to shower leakage from the ECL.
This is taken into account by an ansatz defined by the product of two Crystal Ball~\cite{CrystalBall} functions, given as C($x,\mu,\sigma,\alpha,n$) below.  
\begin{eqnarray}
\label{eq:CB}
  \text{Define: } \: y & = & \frac{x-\mu}{\sigma} \nonumber \\
  \text{ then for } \: y & \geq  & -|\alpha| \nonumber \\ 
  \text{C}(y) & = &  {\rm e}^{-\frac{1}{2}y^{2}},  \nonumber \\
  \mbox{ while for } \: y & <  & -|\alpha| \nonumber \\ 
    \text{C}(y) & = & \left(\frac{n}{|\alpha|}\right)^{n}{\rm e}^{-\frac{1}{2}\alpha^{2}}\left(\frac{n}{|\alpha|} - |\alpha|-y\right)^{-n}.  
     \end{eqnarray}
Here, $\mu$ and $\sigma$ are the mean and width of the Gaussian core, while $\alpha$ and $n$ describe the tail to the lower side of the function. We describe the correlated PDF as:
\begin{equation} \label{eq:mbc_DE}
  P^{s}(M_{\text{bc}},\Delta E) =   \text{C}_{\Delta E}(\Delta E, M_{\text{bc}})
  \text{C}_{M_{\text{bc}}}(M_{\text{bc}}, \Delta E  ).
\end{equation}
In this formulation, $ \text{C}_{\Delta E}(\Delta E, M_{\text{bc}})$ describes the $\Delta E$ shape but has an $M_{\text{bc}}$ dependence and vice versa.
The mean of  C$_{\Delta E}$ has a Gaussian dependence on $M_{\text{bc}}$, while for C$_{M_{\text{bc}}}$, the mean and width both have a linear dependence on $\Delta E$ and the $\alpha$ parameter has a Gaussian dependence on $\Delta E$.
\begin{eqnarray} \label{eq:CB_1}
  \mbox{Define: } \mu^{\prime}_{\text{bc}} & = & \mu_{\text{bc}} + A\times\Delta E \nonumber \\
  \Delta E^{\prime} & = & \Delta E + B\times {\rm e}^{-\frac{1}{2}\left(\frac{M_{\text{bc}} - \mu^{\prime}_{\text{bc}}}{D}\right)^{2}} \nonumber \\
 \text{C}_{\Delta E}(\Delta E, M_{\text{bc}}) & = & \text{C}(\Delta E^{'}, \mu_{\Delta E},\sigma_{\Delta E},\alpha_{\Delta E},n_{\Delta E}) \nonumber\\
 \text{then define: } \mu^{\prime\prime}_{bc} & = & F + G\times \Delta E \nonumber \\
 \sigma^{\prime}_{bc} & = & H + I \times \Delta E \nonumber \\
 \alpha^{\prime}_{bc} & = & J + K \times {\rm e}^{-\frac{1}{2}\left(\frac{\Delta E}{L}\right)^{2}} \nonumber \\
 \text{C}_{M_{\text{bc}}}(M_{\text{bc}},\Delta E) & = & C(M_{\rm bc},\mu^{\prime\prime}_{\rm bc},\sigma^{\prime}_{\rm bc},\alpha^{\prime}_{\rm bc},n_{\rm bc}). 
\end{eqnarray}
Here, $\mu_{\text{\rm bc}}$, $A$, $B$, $D$, $\mu_{\Delta E}$, $\sigma_{\Delta E}$, $\alpha_{\Delta E}$, $n_{\Delta E}$, $F$--$L$, and $n_{\rm bc}$ are the 16 parameters of the correlated function.
No correlation of $T_{c}$ with $M_{\text{bc}}$ or $\Delta E$ is observed.
We model the signal PDF dependence on $T_{c}$ with the sum of a beta distribution~\cite{beta_dist}, a triple Gaussian and a fifth-order polynomial.
\begin{eqnarray} \label{eq:Tc_sig}
  P^{s}(T_{c}) & = & f_{\beta}\beta(T_{c},\beta_{a},\beta_{b}) + \sum_{j=1}^{3}g_{j}{\rm e}^{-\frac{1}{2}\left(\frac{T_{c}- \mu_{j}}{\sigma_{j}}\right)^2}  \nonumber \\
  & + & \sum_{i = 1}^{5}a_{i}T_{c}^{i}.
\end{eqnarray}
Here, $\beta(T_{c},\beta_{a},\beta_{b})$ is the beta distribution, and $f_{\beta}$, $\beta_{a}$, $\beta_{b}$, $g_{j}$, $\mu_{j}$, $\sigma_{j}$, and $a_{i}$ are constants employed in the parameterization. 

The PDFs for the $\rho\pi$ and rare backgrounds are the product of an ARGUS function~\cite{argus} in $M_{\text{bc}}$ and a Crystal Ball function in $\Delta E$.
For $T_{c}$ we employ the same function as the signal PDF shown in Eq.~(\ref{eq:Tc_sig}).
The PDF for the continuum background ($P^{c}$) is the product of an ARGUS function in $M_{\text{bc}}$, a second-order polynomial in $\Delta E$, and a seventh-order polynomial in $T_{c}$ that is constrained to be greater than zero.
For each of $P^{s}$, $P^{c}$, $P^{\rho \pi}$ and $P^{r}$ we find no dependence on $q\cdot r$ for the $M_{\text{bc}}$ and $\Delta E$ variables.
Consequently the parameterization of these PDFs as a function of $M_{\text{bc}}$ and $\Delta E$ do not vary in bins of $q\cdot r$.
For the $T_{c}$ dependence, the PDF distributions are fit for each bin in $q\cdot r$ to account for an observed dependence on this variable.
In the case of $P^{c}$, the parameters for the $M_{\text{bc}}$ and $\Delta E$ variables are the same for all bins in $q\cdot r$ and are free to float in the fit.
The parameters for its $T_{c}$ dependence are determined from off-resonance data and fixed in the final fit.

All PDFs and their products are properly normalized. 
The PDF shape parameters for signal, $\rho\pi$, and rare backgrounds are determined from fits to large samples of MC events and fixed in the final fit.
In total, there are 16 free parameters in the fit, including the parameters of $M_{\text{bc}}$ and $\Delta E$ components of the continuum PDF. All other parameters are fixed. 


The systematic uncertainties introduced by the above assumptions for  $P^{c}$ are determined from MC simulations of the continuum background.
To test the assumption that for the $T_{c}$ dependence of $P^{c}$ one can employ off-resonance data to model the on-resonance continuum, we first build a model of signal plus backgrounds and determine the $P^{c}$ parameterization by a fit to MC simulations of the off-resonance data.
We compare the signal yield extracted from this off-resonance parameterization to that extracted when the parameterization is determined by fits to the signal region of MC simulations.
These simulations are equivalent to six times the data recorded by the experiment.
To test the assumption that a single parameterization of the $M_{\text{bc}}$ and $\Delta E$ dependence of the $P^{c}$ PDF can be used for all bins in $q\cdot r$, we fit $P^{c}(M_{\text{bc}}, \Delta E$) to off-resonance data in bins of $q\cdot r$.
These parameterizations are used to generate toy MC events which are fitted with a single $P^{c}(M_{\text{bc}}, \Delta E$)  for all bins in $q\cdot r$.
The differences in yield from these studies are used to determine the systematic uncertainties.

The fitting procedure and fidelity of the various PDF models are extensively investigated in toy MC studies.
In these, the signal, $\rho\pi$, and rare background events are selected from large samples of simulated events.
Events for $e^{+} e^{-} \to q \bar{q}$  are generated from the continuum PDF shapes.
We observe a $1\%$ ($2\%$) bias for the yield ($A_{\CP}$) due to limitations of the PDF ansatzes used to model the data.
This bias is included as a systematic error in the final $\mathcal{B}$ and $A_{\CP}$ calculation. 
A high-statistics sample of $\tau^{+} \to \pi^{+} \pi^{0} \nu_{\tau}$ decays~\cite{Ryu} is used to correct the prediction for the efficiency of $\pi^{0}$ detection.

\begin{figure*}[htb]
  \includegraphics[width=0.78\textwidth]{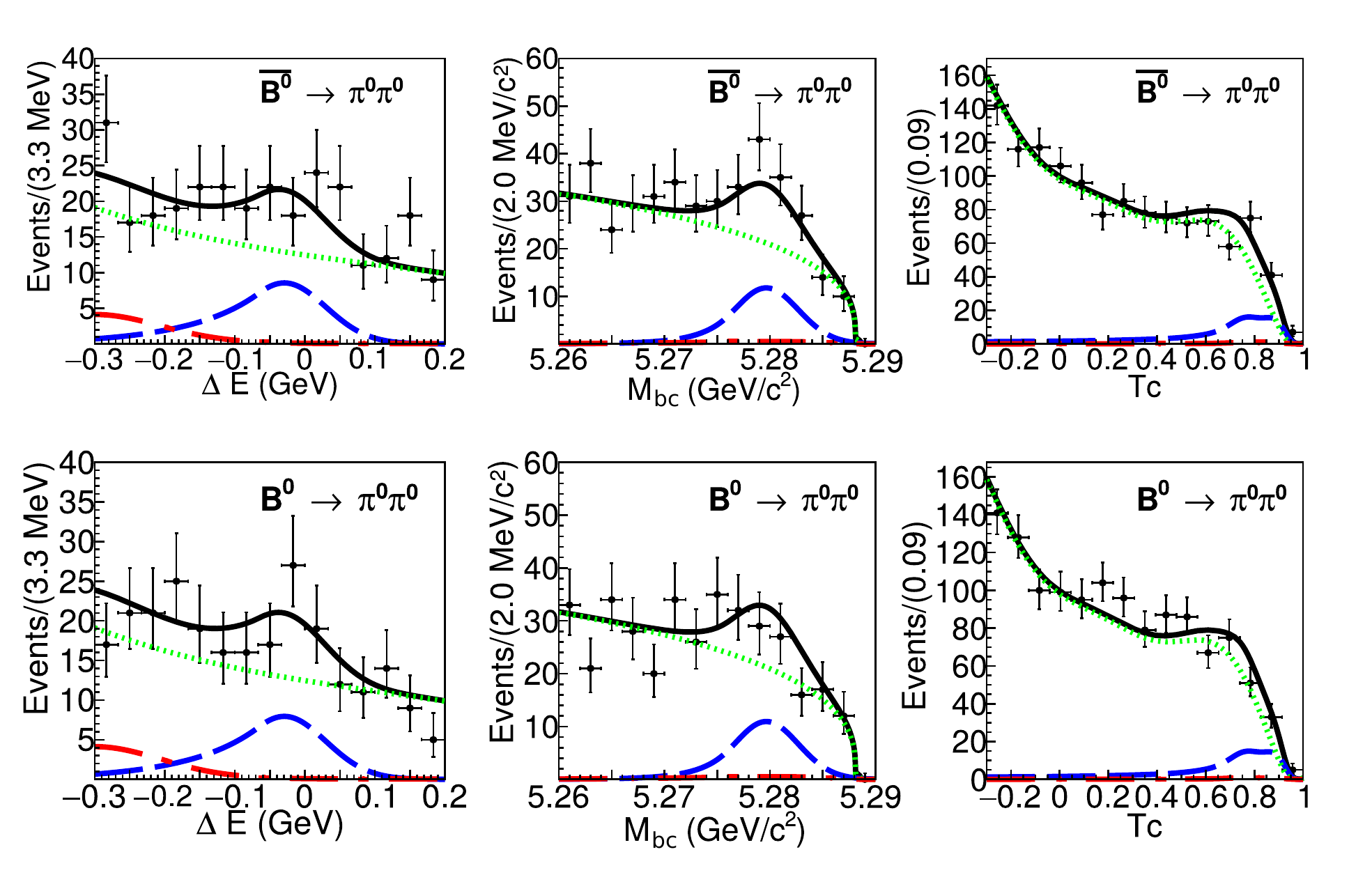}
\caption{Projections of the fit results onto (left) $\Delta E$, (middle) $M_{\text{bc}}$, (right) $T_{c}$ are shown in the signal enhanced region: $5.275\gevcc < M_{\text{bc}} < 5.285\gevcc$, $-0.15\gev < \Delta E < 0.05\gev$, and $T_{c} > 0.7$. Each panel shows the distribution enhanced in the other two variables.
Data are points with error bars, and fit results are shown by the solid black curves. Contributions from signal, continuum $q\bar{q}$, combined $\rho \pi$ and other rare $B$ decays are shown by the dashed blue, dotted green, and dash-dotted red curves, respectively. The top (bottom) row panels are for events with positive (negative) $q$ tags.}
\label{yield}
\end{figure*}

Figure~\ref{yield} shows the signal-enhanced projections of the fits to data in  $M_{\text{bc}}$, $\Delta E$ and $T_{c}$.
We obtain a signal yield of $217 \pm 32$ events.
Assuming the $\Upsilon(4S)$ decays to charged and neutral $B$ modes equally, and a final detection efficiency after all selections and corrections of $22\%$, we determine the branching fraction to be
\begin{equation}
\label{eq:BR}
 {\cal B} (B^{0} \to  \pi^{0}\pi^{0}) = (1.31 \pm 0.19 \pm 0.19) \times 10^{-6}, 
\end{equation}
where the quoted uncertainties are statistical and systematic, respectively.
The systematic uncertainties include contributions due to the continuum background parameterization in $T_{c}$ ($11.0\%$), $\pi^{0}$ detection efficiency ($4.4\%$), single continuum parameterization for $M_{\text{bc}}$ and $\Delta E$ ($4.0\%$), assumed $\mathcal{B}$ for $B^{+} \to \rho^{+} \pi^{0}$ ($4.0\%$), off-resonance continuum background ($3.0\%$), assumed $\mathcal{B}$ for other rare decays ($3.0\%$),  determination of $f^{c}_{i,d}$ fraction($1.8\%$), the choice of fitted region ($1.5\%$),  $f^{\rho \pi}_{i,d}$ and $f^{r}_{i,d}$ fractions equal to $f^{s}_{i,d}$ ($1.5\%$), luminosity (including assumption of equal branching fraction for charged and neutral modes) ($1.4\%$), fit bias ($1.0\%$), recovery of converted photons ($1.0\%$), and timing cut ($0.5\%$).  Adding these in quadrature gives a total systematic uncertainty of $14.2\%$. 

The significance of the result is determined by convolving the statistical and additive systematic uncertainties and calculating $\sqrt{2({\cal L}_{m} - {\cal L}_{0})}$, where ${\cal L}_{m}$ is the log-likelihood for the measured yield and ${\cal L}_{0}$ is that for a null yield. This gives a total significance of $6.4$ standard deviations.
The direct $\CP$ violation parameter is measured to be
\begin{equation} \label{eq:Acp}
A_{\CP} = +0.14 \pm 0.36 \pm 0.10.
\end{equation}
The second uncertainty is systematic, and is the quadratic sum of possible effects on $A_{\CP}$ of uncertainties in the continuum background parameterization of $T_{c}$ (0.08), $\rho\pi$ and other rare backgrounds (0.06), and fit bias (0.02).

As a cross-check, a separate flavor-independent analysis is performed employing an artificial neural network in lieu of $T_{c}$ for continuum suppression. Though this analysis has $1\%$ less signal efficiency, the measured branching fraction agrees with the flavor-dependent measurement within uncertainties.

\begin{figure}[htb]
\includegraphics[width=0.47\textwidth]{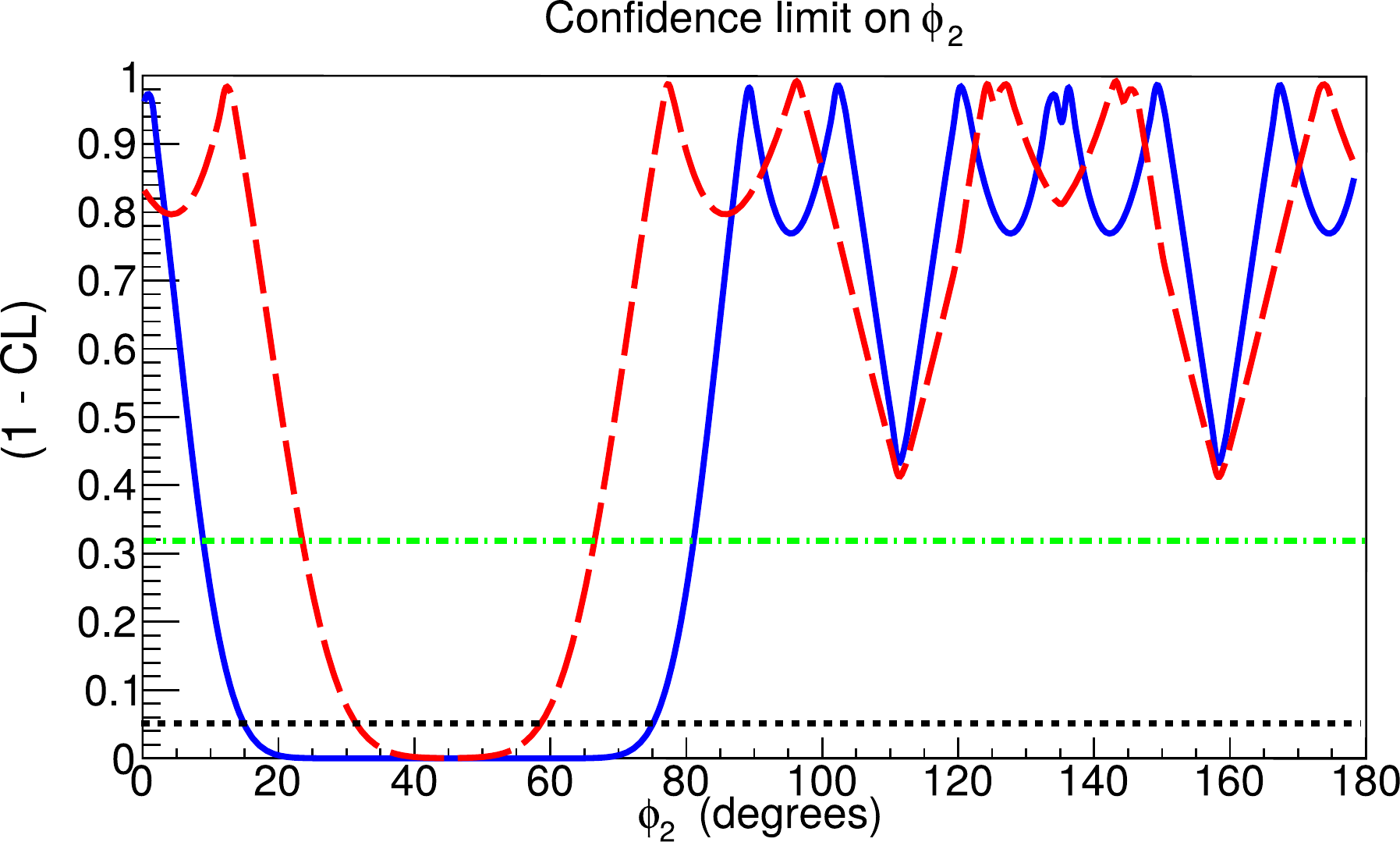}
\caption{Scan of the confidence level for $\phi_{2}$ using only data from $B\to \pi \pi$ measurements of the Belle experiment.
The dashed red curve shows the previous constraint from Belle data~\cite{Dalseno}, the solid blue curve includes our new results.
The updated results for $B^{0} \to \pi^{0} \pi^{0}$ exclude  $9.5^{\circ} < \phi_{2} < 81.6^{\circ}$ at the 68\% confidence level (grean dot-dashed line) and $15.5^{\circ} < \phi_{2} < 75.0^{\circ}$ at 95\% confidence level (black dashed line).}
\label{isospin}
\end{figure}

Combining our results for the $\mathcal{B}$ and $A_{\CP}$ for $B^{0} \to \pi^{0} \pi^{0}$ with Belle's previous measurements of $\mathcal{B}$ and time-dependent $\CP$ violation for $B^{0} \to \pi^{+} \pi^{-}$~\cite{Dalseno} and $\mathcal{B}$ and $A_{\CP}$ for $B^{+} \to \pi^{+} \pi^{0}$~\cite{Duh} allows us to employ the isopsin analysis of Ref.~\cite{GronauLondon} to constrain $\phi_{2}$.
The result of the fit is shown in Fig.~\ref{isospin}.
Our results exclude  $15.5^{\circ} < \phi_{2} < 75.0^{\circ}$ at 95\% confidence level.

The measured branching fraction is smaller than our previously published result~\cite{Belle_pi0pi0} though consistent within uncertainties.
The difference could be due to a substantially smaller fraction of data for which ECL timing information was available (113 of 253 fb$^{-1}$) in the earlier measurement and the subsequent extrapolation to the full data set.
The result reported here supersedes our earlier published values and agrees with BaBar measurement~\cite{Babar_hh} within combined uncertainties.
While this result is closer to theory predictions than the earlier Belle~\cite{Belle_pi0pi0} and BaBar~\cite{Babar_hh} measurements, it is still larger than expectations based on the factorization model~\cite{Factorization}.
It is in agreement with the recent works of Qiao {\it et al}.~\cite{conformality} as well as Li and Yu~\cite{Pertubative} which employ different theoretical approaches.
The upcoming Belle II experiment~\cite{belleii}, with its projected factor of 50 increase in luminosity, will enable precision measurements of $\mathcal{B}$ and $\CP$ asymmetry of $B^{0} \to \pi^{0} \pi^{0}$ and other $B\to \pi\pi$ decays to strongly constrain $\phi_{2}$.

%




%
%

We thank the KEKB group for the excellent operation of the
accelerator; the KEK cryogenics group for the efficient
operation of the solenoid; and the KEK computer group,
the National Institute of Informatics, and the 
PNNL/EMSL computing group for valuable computing
and SINET5 network support.  We acknowledge support from
the Ministry of Education, Culture, Sports, Science, and
Technology (MEXT) of Japan, the Japan Society for the 
Promotion of Science (JSPS), and the Tau-Lepton Physics 
Research Center of Nagoya University; 
the Australian Research Council;
Austrian Science Fund under Grant No.~P 26794-N20;
the National Natural Science Foundation of China under Contracts 
No.~10575109, No.~10775142, No.~10875115, No.~11175187, No.~11475187, 
No.~11521505 and No.~11575017;
the Chinese Academy of Science Center for Excellence in Particle Physics; 
the Ministry of Education, Youth and Sports of the Czech
Republic under Contract No.~LTT17020;
the Carl Zeiss Foundation, the Deutsche Forschungsgemeinschaft, the
Excellence Cluster Universe, and the VolkswagenStiftung;
the Department of Science and Technology of India; 
the Istituto Nazionale di Fisica Nucleare of Italy; 
the WCU program of the Ministry of Education, National Research Foundation (NRF)
of Korea Grants No.~2011-0029457, No.~2012-0008143,
No.~2014R1A2A2A01005286,
No.~2014R1A2A2A01002734, No.~2015R1A2A2A01003280,
No.~2015H1A2A1033649, No.~2016R1D1A1B01010135, No.~2016K1A3A7A09005603, No.~2016K1A3A7A09005604, No.~2016R1D1A1B02012900,
No.~2016K1A3A7A09005606, No.~NRF-2013K1A3A7A06056592;
the Brain Korea 21-Plus program, Radiation Science Research Institute, Foreign Large-size Research Facility Application Supporting project and the Global Science Experimental Data Hub Center of the Korea Institute of Science and Technology Information;
the Polish Ministry of Science and Higher Education and 
the National Science Center;
the Ministry of Education and Science of the Russian Federation and
the Russian Foundation for Basic Research;
the Slovenian Research Agency;
Ikerbasque, Basque Foundation for Science and
MINECO (Juan de la Cierva), Spain;
the Swiss National Science Foundation; 
the Ministry of Education and the Ministry of Science and Technology of Taiwan;
and the U.S.\ Department of Energy and the National Science Foundation.

\bibliography{PRD_B_pi0pi0_submitted}

\begin{thebibliography}{29}%
\makeatletter
\providecommand \@ifxundefined [1]{%
 \@ifx{#1\undefined}
}%
\providecommand \@ifnum [1]{%
 \ifnum #1\expandafter \@firstoftwo
 \else \expandafter \@secondoftwo
 \fi
}%
\providecommand \@ifx [1]{%
 \ifx #1\expandafter \@firstoftwo
 \else \expandafter \@secondoftwo
 \fi
}%
\providecommand \natexlab [1]{#1}%
\providecommand \enquote  [1]{``#1''}%
\providecommand \bibnamefont  [1]{#1}%
\providecommand \bibfnamefont [1]{#1}%
\providecommand \citenamefont [1]{#1}%
\providecommand \href@noop [0]{\@secondoftwo}%
\providecommand \href [0]{\begingroup \@sanitize@url \@href}%
\providecommand \@href[1]{\@@startlink{#1}\@@href}%
\providecommand \@@href[1]{\endgroup#1\@@endlink}%
\providecommand \@sanitize@url [0]{\catcode `\\12\catcode `\$12\catcode
  `\&12\catcode `\#12\catcode `\^12\catcode `\_12\catcode `\%12\relax}%
\providecommand \@@startlink[1]{}%
\providecommand \@@endlink[0]{}%
\providecommand \url  [0]{\begingroup\@sanitize@url \@url }%
\providecommand \@url [1]{\endgroup\@href {#1}{\urlprefix }}%
\providecommand \urlprefix  [0]{URL }%
\providecommand \Eprint [0]{\href }%
\providecommand \doibase [0]{http://dx.doi.org/}%
\providecommand \selectlanguage [0]{\@gobble}%
\providecommand \bibinfo  [0]{\@secondoftwo}%
\providecommand \bibfield  [0]{\@secondoftwo}%
\providecommand \translation [1]{[#1]}%
\providecommand \BibitemOpen [0]{}%
\providecommand \bibitemStop [0]{}%
\providecommand \bibitemNoStop [0]{.\EOS\space}%
\providecommand \EOS [0]{\spacefactor3000\relax}%
\providecommand \BibitemShut  [1]{\csname bibitem#1\endcsname}%
\let\auto@bib@innerbib\@empty
\bibitem [{\citenamefont {Bevan}\ \emph {et~al.}(2014)\citenamefont {Bevan}
  \emph {et~al.}}]{PBF}%
  \BibitemOpen
  \bibfield  {author} {\bibinfo {author} {\bibfnamefont {A.}~\bibnamefont
  {Bevan}} \emph {et~al.},\ }\href@noop {} {\bibfield  {journal} {\bibinfo
  {journal} {Eur. Phys. J. C}\ }\textbf {\bibinfo {volume} {74}},\ \bibinfo
  {pages} {3026} (\bibinfo {year} {2014})}\BibitemShut {NoStop}%
\bibitem [{\citenamefont {Aaij}\ \emph {et~al.}(2015)\citenamefont {Aaij} \emph
  {et~al.}}]{LHCb_CPV_phi1}%
  \BibitemOpen
  \bibfield  {author} {\bibinfo {author} {\bibfnamefont {R.}~\bibnamefont
  {Aaij}} \emph {et~al.} (\bibinfo {collaboration} {LHCb Collaboration}),\
  }\href@noop {} {\bibfield  {journal} {\bibinfo  {journal} {Phys. Rev. Lett.}\
  }\textbf {\bibinfo {volume} {115}},\ \bibinfo {pages} {031601} (\bibinfo
  {year} {2015})}\BibitemShut {NoStop}%
\bibitem [{\citenamefont {Aaij}\ \emph {et~al.}(2016)\citenamefont {Aaij} \emph
  {et~al.}}]{LHCb_CPV_phi3}%
  \BibitemOpen
  \bibfield  {author} {\bibinfo {author} {\bibfnamefont {R.}~\bibnamefont
  {Aaij}} \emph {et~al.} (\bibinfo {collaboration} {LHCb Collaboration}),\
  }\href@noop {} {\bibfield  {journal} {\bibinfo  {journal} {JHEP}\ }\textbf
  {\bibinfo {volume} {2016}},\ \bibinfo {pages} {87} (\bibinfo {year}
  {2016})}\BibitemShut {NoStop}%
\bibitem [{\citenamefont {Kobayashi}\ and\ \citenamefont {Maskawa}(1973)}]{KM}%
  \BibitemOpen
  \bibfield  {author} {\bibinfo {author} {\bibfnamefont {M.}~\bibnamefont
  {Kobayashi}}\ and\ \bibinfo {author} {\bibfnamefont {T.}~\bibnamefont
  {Maskawa}},\ }\href@noop {} {\bibfield  {journal} {\bibinfo  {journal} {Prog.
  Theor. Phys.}\ }\textbf {\bibinfo {volume} {49}},\ \bibinfo {pages} {652}
  (\bibinfo {year} {1973})}\BibitemShut {NoStop}%
\bibitem [{alp()}]{alpha}%
  \BibitemOpen
  \href@noop {} {}\bibinfo {note} {Another naming convention, $\beta$ ($=
  \phi_1$), $\alpha$ ($= \phi_2$) and $\gamma$ ($=\phi_3$) is also used in the
  literature.}\BibitemShut {Stop}%
\bibitem [{\citenamefont {Gronau}\ and\ \citenamefont
  {London}(1990)}]{GronauLondon}%
  \BibitemOpen
  \bibfield  {author} {\bibinfo {author} {\bibfnamefont {M.}~\bibnamefont
  {Gronau}}\ and\ \bibinfo {author} {\bibfnamefont {D.}~\bibnamefont
  {London}},\ }\href@noop {} {\bibfield  {journal} {\bibinfo  {journal} {Phys.
  Rev. Lett.}\ }\textbf {\bibinfo {volume} {65}},\ \bibinfo {pages} {3381}
  (\bibinfo {year} {1990})}\BibitemShut {NoStop}%
\bibitem [{\citenamefont {Dalseno}\ \emph {et~al.}(2013)\citenamefont {Dalseno}
  \emph {et~al.}}]{Dalseno}%
  \BibitemOpen
  \bibfield  {author} {\bibinfo {author} {\bibfnamefont {J.}~\bibnamefont
  {Dalseno}} \emph {et~al.} (\bibinfo {collaboration} {Belle Collaboration}),\
  }\href@noop {} {\bibfield  {journal} {\bibinfo  {journal} {Phys. Rev. D}\
  }\textbf {\bibinfo {volume} {88}},\ \bibinfo {pages} {092003} (\bibinfo
  {year} {2013})}\BibitemShut {NoStop}%
\bibitem [{CC()}]{CC}%
  \BibitemOpen
  \href@noop {} {}\bibinfo {note} {Throughout this Letter, the inclusion of the
  charge-conjugate decay modes is implied unless otherwise stated.}\BibitemShut
  {Stop}%
\bibitem [{\citenamefont {Li}\ and\ \citenamefont
  {Mishima}(2006)}]{LiMishima1}%
  \BibitemOpen
  \bibfield  {author} {\bibinfo {author} {\bibfnamefont {H.}~\bibnamefont
  {Li}}\ and\ \bibinfo {author} {\bibfnamefont {S.}~\bibnamefont {Mishima}},\
  }\href@noop {} {\bibfield  {journal} {\bibinfo  {journal} {Phys. Rev. D}\
  }\textbf {\bibinfo {volume} {73}},\ \bibinfo {pages} {114014} (\bibinfo
  {year} {2006})}\BibitemShut {NoStop}%
\bibitem [{\citenamefont {Li}\ and\ \citenamefont
  {Mishima}(2011)}]{LiMishima2}%
  \BibitemOpen
  \bibfield  {author} {\bibinfo {author} {\bibfnamefont {H.}~\bibnamefont
  {Li}}\ and\ \bibinfo {author} {\bibfnamefont {S.}~\bibnamefont {Mishima}},\
  }\href@noop {} {\bibfield  {journal} {\bibinfo  {journal} {Phys. Rev. D}\
  }\textbf {\bibinfo {volume} {83}},\ \bibinfo {pages} {034023} (\bibinfo
  {year} {2011})}\BibitemShut {NoStop}%
\bibitem [{\citenamefont {Chao}\ \emph {et~al.}(2005)\citenamefont {Chao} \emph
  {et~al.}}]{Belle_pi0pi0}%
  \BibitemOpen
  \bibfield  {author} {\bibinfo {author} {\bibfnamefont {Y.}~\bibnamefont
  {Chao}} \emph {et~al.} (\bibinfo {collaboration} {Belle Collaboration}),\
  }\href@noop {} {\bibfield  {journal} {\bibinfo  {journal} {Phys. Rev. Lett.}\
  }\textbf {\bibinfo {volume} {94}},\ \bibinfo {pages} {181803} (\bibinfo
  {year} {2005})}\BibitemShut {NoStop}%
\bibitem [{\citenamefont {Lees}\ \emph {et~al.}(2013)\citenamefont {Lees} \emph
  {et~al.}}]{Babar_hh}%
  \BibitemOpen
  \bibfield  {author} {\bibinfo {author} {\bibfnamefont {J.}~\bibnamefont
  {Lees}} \emph {et~al.} (\bibinfo {collaboration} {BaBar Collaboration}),\
  }\href@noop {} {\bibfield  {journal} {\bibinfo  {journal} {Phys. Rev. D}\
  }\textbf {\bibinfo {volume} {87}},\ \bibinfo {pages} {052009} (\bibinfo
  {year} {2013})}\BibitemShut {NoStop}%
\bibitem [{KEK()}]{KEKB}%
  \BibitemOpen
  \href@noop {} {}\bibinfo {note} {S. Kurokawa and E. Kikutani, Nucl. Instrum.
  Methods Phys. Res., Sect. A {\bf 499}, 1 (2003), and other papers included in
  this Volume; T. Abe {\it et al.}, Prog. Theor. Exp. Phys. (2013) 03A001 and
  following articles up to 03A011.}\BibitemShut {Stop}%
\bibitem [{Bel()}]{Belle}%
  \BibitemOpen
  \href@noop {} {}\bibinfo {note} {A. Abashian {\it et al.} (Belle
  Collaboration), Nucl. Instrum. Methods Phys. Res., Sect. A {\bf 479}, 117
  (2002); also see the detector section in J. Brodzicka {\it et al.}, Prog.
  Theor. Exp. Phys. (2012) 04D001.}\BibitemShut {Stop}%
\bibitem [{\citenamefont {Natkaniec}\ \emph {et~al.}(2006)\citenamefont
  {Natkaniec} \emph {et~al.}}]{svd2}%
  \BibitemOpen
  \bibfield  {author} {\bibinfo {author} {\bibfnamefont {Z.}~\bibnamefont
  {Natkaniec}} \emph {et~al.} (\bibinfo {collaboration} {Belle SVD2 Group}),\
  }\href@noop {} {\bibfield  {journal} {\bibinfo  {journal} {Nucl. Instrum.
  Methods Phys. Res., Sect. A}\ }\textbf {\bibinfo {volume} {560}},\ \bibinfo
  {pages} {1} (\bibinfo {year} {2006})}\BibitemShut {NoStop}%
\bibitem [{\citenamefont {Patrignani}\ \emph {et~al.}(2016)\citenamefont
  {Patrignani} \emph {et~al.}}]{PDG_chi_d}%
  \BibitemOpen
  \bibfield  {author} {\bibinfo {author} {\bibfnamefont {C.}~\bibnamefont
  {Patrignani}} \emph {et~al.} (\bibinfo {collaboration} {Particle Data
  Group}),\ }\href@noop {} {\bibfield  {journal} {\bibinfo  {journal} {Chin.
  Phys. C}\ }\textbf {\bibinfo {volume} {40}},\ \bibinfo {pages} {100001}
  (\bibinfo {year} {2016})}\BibitemShut {NoStop}%
\bibitem [{SFW()}]{SFW}%
  \BibitemOpen
  \href@noop {} {}\bibinfo {note} {The Fox-Wolfram moments were introduced by
  G.~C.~Fox and S.~Wolfram in Phys.\ Rev.\ Lett.\ {\bf 41}, 1581 (1978). The
  modified Fox-Wolfram moments used in this paper are described in S.~H.~Lee
  {\it et al.} (Belle Collaboration), Phys.\ Rev.\ Lett.\ {\bf 91}, 261801
  (2003).}\BibitemShut {Stop}%
\bibitem [{\citenamefont {Lange}\ \emph {et~al.}(2001)\citenamefont {Lange}
  \emph {et~al.}}]{evtgen}%
  \BibitemOpen
  \bibfield  {author} {\bibinfo {author} {\bibfnamefont {D.}~\bibnamefont
  {Lange}} \emph {et~al.},\ }\href@noop {} {\bibfield  {journal} {\bibinfo
  {journal} {Nucl. Instrum. Methods Phys. Res., Sect. A}\ }\textbf {\bibinfo
  {volume} {462}},\ \bibinfo {pages} {152} (\bibinfo {year}
  {2001})}\BibitemShut {NoStop}%
\bibitem [{\citenamefont {Brun}\ \emph {et~al.}(1987)\citenamefont {Brun} \emph
  {et~al.}}]{GEANT}%
  \BibitemOpen
  \bibfield  {author} {\bibinfo {author} {\bibfnamefont {R.}~\bibnamefont
  {Brun}} \emph {et~al.},\ }\href@noop {} {\bibfield  {journal} {\bibinfo
  {journal} {CERN Report No. DD/EE/84-1}\ } (\bibinfo {year}
  {1987})}\BibitemShut {NoStop}%
\bibitem [{\citenamefont {Kakuno}\ \emph {et~al.}(2004)\citenamefont {Kakuno}
  \emph {et~al.}}]{kukuno}%
  \BibitemOpen
  \bibfield  {author} {\bibinfo {author} {\bibfnamefont {H.}~\bibnamefont
  {Kakuno}} \emph {et~al.},\ }\href@noop {} {\bibfield  {journal} {\bibinfo
  {journal} {Nucl. Instrum. Methods Phys. Res., Sect. A}\ }\textbf {\bibinfo
  {volume} {533}},\ \bibinfo {pages} {516} (\bibinfo {year}
  {2004})}\BibitemShut {NoStop}%
\bibitem [{Cry()}]{CrystalBall}%
  \BibitemOpen
  \href@noop {} {}\bibinfo {note} {T. Skwarnicki, DESY F31-86-02, (1986)
  (unpublished)}\BibitemShut {NoStop}%
\bibitem [{\citenamefont {Fente}\ \emph {et~al.}(1999)\citenamefont {Fente},
  \citenamefont {Knutson},\ and\ \citenamefont {Schexnayder}}]{beta_dist}%
  \BibitemOpen
  \bibfield  {author} {\bibinfo {author} {\bibfnamefont {J.}~\bibnamefont
  {Fente}}, \bibinfo {author} {\bibfnamefont {K.}~\bibnamefont {Knutson}}, \
  and\ \bibinfo {author} {\bibfnamefont {C.}~\bibnamefont {Schexnayder}},\ }in\
  \href@noop {} {\emph {\bibinfo {booktitle} {Proceedings of the 31st
  conference on Winter simulation: Simulation---a bridge to the future-Volume
  2}}}\ (\bibinfo {organization} {ACM},\ \bibinfo {year} {1999})\ pp.\ \bibinfo
  {pages} {1010--1015}\BibitemShut {NoStop}%
\bibitem [{\citenamefont {Albrecht}\ \emph {et~al.}(1990)\citenamefont
  {Albrecht} \emph {et~al.}}]{argus}%
  \BibitemOpen
  \bibfield  {author} {\bibinfo {author} {\bibfnamefont {H.}~\bibnamefont
  {Albrecht}} \emph {et~al.} (\bibinfo {collaboration} {ARGUS Collaboration}),\
  }\href@noop {} {\bibfield  {journal} {\bibinfo  {journal} {Phys. Lett. B}\
  }\textbf {\bibinfo {volume} {241}},\ \bibinfo {pages} {278} (\bibinfo {year}
  {1990})}\BibitemShut {NoStop}%
\bibitem [{\citenamefont {Ryu}\ \emph {et~al.}(2014)\citenamefont {Ryu} \emph
  {et~al.}}]{Ryu}%
  \BibitemOpen
  \bibfield  {author} {\bibinfo {author} {\bibfnamefont {S.}~\bibnamefont
  {Ryu}} \emph {et~al.} (\bibinfo {collaboration} {Belle Collaboration}),\
  }\href@noop {} {\bibfield  {journal} {\bibinfo  {journal} {Phys. Rev. D}\
  }\textbf {\bibinfo {volume} {89}},\ \bibinfo {pages} {072009} (\bibinfo
  {year} {2014})}\BibitemShut {NoStop}%
\bibitem [{\citenamefont {Duh}\ \emph {et~al.}(2012)\citenamefont {Duh} \emph
  {et~al.}}]{Duh}%
  \BibitemOpen
  \bibfield  {author} {\bibinfo {author} {\bibfnamefont {Y.-T.}\ \bibnamefont
  {Duh}} \emph {et~al.} (\bibinfo {collaboration} {Belle Collaboration}),\
  }\href@noop {} {\bibfield  {journal} {\bibinfo  {journal} {Phys. Rev. D}\
  }\textbf {\bibinfo {volume} {87}},\ \bibinfo {pages} {031103(R)} (\bibinfo
  {year} {2012})}\BibitemShut {NoStop}%
\bibitem [{\citenamefont {Zhang}\ \emph {et~al.}(2014)\citenamefont {Zhang},
  \citenamefont {Liu}, \citenamefont {Fan}, \citenamefont {Cheng},\ and\
  \citenamefont {Xiao}}]{Factorization}%
  \BibitemOpen
  \bibfield  {author} {\bibinfo {author} {\bibfnamefont {Y.-L.}\ \bibnamefont
  {Zhang}}, \bibinfo {author} {\bibfnamefont {X.-Y.}\ \bibnamefont {Liu}},
  \bibinfo {author} {\bibfnamefont {Y.-Y.}\ \bibnamefont {Fan}}, \bibinfo
  {author} {\bibfnamefont {S.}~\bibnamefont {Cheng}}, \ and\ \bibinfo {author}
  {\bibfnamefont {Z.-J.}\ \bibnamefont {Xiao}},\ }\href@noop {} {\bibfield
  {journal} {\bibinfo  {journal} {Phys. Rev. D}\ }\textbf {\bibinfo {volume}
  {90}},\ \bibinfo {pages} {014029} (\bibinfo {year} {2014})}\BibitemShut
  {NoStop}%
\bibitem [{\citenamefont {Qiao}\ \emph {et~al.}(2015)\citenamefont {Qiao},
  \citenamefont {Zhu}, \citenamefont {Wu},\ and\ \citenamefont
  {Brodsky}}]{conformality}%
  \BibitemOpen
  \bibfield  {author} {\bibinfo {author} {\bibfnamefont {C.-F.}\ \bibnamefont
  {Qiao}}, \bibinfo {author} {\bibfnamefont {R.-L.}\ \bibnamefont {Zhu}},
  \bibinfo {author} {\bibfnamefont {X.-G.}\ \bibnamefont {Wu}}, \ and\ \bibinfo
  {author} {\bibfnamefont {S.~J.}\ \bibnamefont {Brodsky}},\ }\href@noop {}
  {\bibfield  {journal} {\bibinfo  {journal} {Phys. Lett. B}\ }\textbf
  {\bibinfo {volume} {748}},\ \bibinfo {pages} {422 } (\bibinfo {year}
  {2015})}\BibitemShut {NoStop}%
\bibitem [{\citenamefont {Li}\ and\ \citenamefont {Yu}(2017)}]{Pertubative}%
  \BibitemOpen
  \bibfield  {author} {\bibinfo {author} {\bibfnamefont {Y.-F.}\ \bibnamefont
  {Li}}\ and\ \bibinfo {author} {\bibfnamefont {X.-Q.}\ \bibnamefont {Yu}},\
  }\href@noop {} {\bibfield  {journal} {\bibinfo  {journal} {Phys. Rev. D}\
  }\textbf {\bibinfo {volume} {95}},\ \bibinfo {pages} {034023} (\bibinfo
  {year} {2017})}\BibitemShut {NoStop}%
\bibitem [{bel()}]{belleii}%
  \BibitemOpen
  \href@noop {} {}\bibinfo {note} {T. Abe {\it et al.}, arXiv:1011.0352
  [physics.ins-det]}\BibitemShut {NoStop}%
\end{thebibliography}%

\end{document}